\documentclass[11pt]{article}
\usepackage[latin1]{inputenc}
\usepackage[english]{babel}
\usepackage[namelimits]{amsmath}
\usepackage{amssymb}
\usepackage{amsmath}
\usepackage{amsthm}

\begin{document}

\title{Vacuum Solutions of Einstein's Equations in Parabolic Coordinates}

\author{Stefano Viaggiu
\\
Dipartimento di Matematica,\\
Universit\'a di Roma ``Tor Vergata'',\\
Via della Ricerca Scientifica, 1, I-00133 Roma, Italy\\
E-mail: viaggiu@mat.uniroma2.it\\
(or: stefano.viaggiu@ax0rm1.roma1.infn.it)}
\date{\today}\maketitle

\begin{abstract}
We present a simple method to obtain vacuum solutions
of Einstein's equations in parabolic
coordinates starting from ones with cylindrical symmetries.
Furthermore, a generalization of the method to a more general
situation is given together with a
discussion of the possible relations between our method and the
Belinsky-Zakharov solitons-generating solutions.
\end{abstract}
\section*{Introduction}
\noindent Cylindrical solutions \cite{Lev,Ch,Lew,Lin,Sant} 
have played an important role in the history of general relativity. In
astrophysics, cylindrical solutions have been
applied in the context of cosmic strings \cite{Wil}.
In literature
various techniques exist \cite{Ger,Xant,Eh,Bel,Zac} for generating solutions of
Einstein's equations. In this paper we introduce a simple method for
generating static and stationary 
``parabolic'' solutions starting from ones with cylindrical symmetries.
Our starting point is the stationary axially symmetric line element 
in the form \cite{Pap} 
\begin{equation}
ds^2=f^{-1}[e^{2\gamma}({(dx^1)}^2+{(dx^2)}^2)+{\rho}^2 d{\phi}^2]-
f{(dt-\omega d\phi)}^2,
\label{A}
\end{equation}
where $x^1, x^2$ are spatial coordinates, $\phi$ is an angular
coordinate, $t$ is a time coordinate,
$\rho$ is the radius in a cylindrical coordinate system
and $f,\gamma,\omega$ are
functions of $x^1,x^2$. The field equations \cite{Ernst}
for the line element
(\ref{A}) can be written \cite{Viag} in the form:
\begin{eqnarray}
& &{\nabla}^2 f-\frac{1}{f}({f_{\alpha}}^2-{{\Phi}_{\alpha}}^2)=0\;,\;
{\nabla}^2\Phi-\frac{2}{f}f_{\alpha}{\Phi}_{\alpha}=0 ,\nonumber\\
& &{\gamma}_1=-\frac{\Sigma{\rho}_1+\Pi{\rho}_2}
{4\rho({\rho}_{1}^{2}+{\rho}_{2}^{2})}+\frac{c}{2}\;\;,\;\;
{\gamma}_2=\frac{\Sigma{\rho}_2-\Pi{\rho}_1}
{4\rho({\rho}_{1}^{2}+{\rho}_{2}^{2})}+\frac{d}{2} ,\nonumber\\
& &c=\frac{2{\rho}_{12}{\rho}_2+({\rho}_{11}-{\rho}_{22}){\rho}_1}
{({\rho}_{1}^{2}+{\rho}_{2}^{2})}\;\;,\;\;
d=\frac{2{\rho}_{12}{\rho}_1-({\rho}_{11}-{\rho}_{22}){\rho}_2}
{({\rho}_{1}^{2}+{\rho}_{2}^2)} ,\nonumber\\
& &\Sigma=\frac{{\rho}^2}{f^2}(f_{2}^2-f_{1}^2)+f^2
({\omega}_{1}^{2}-{\omega}_{2}^2)\;\;,\;\;
\Pi=-2{\rho}^2\frac{f_1 f_2}{f^2}+2f^2{\omega}_1 {\omega}_2 ,\nonumber\\
& &{\omega}_1=-\frac{\rho}{f^2}{\Phi}_2\;\;,\;\;
{\omega}_2=\frac{\rho}{f^2}{\Phi}_1 \label{B}
\end{eqnarray}
where a summation over $\alpha$ is implicit with
$\alpha=1,2$, i.e. $x^1, x^2$, 
and subindices
denote partial derivatives. Further, the operator ${\nabla}^2$,
with 
${\nabla}^2={\partial}^{2}_{\alpha\alpha}+
\frac{{\rho}_{\alpha}}{\rho}{\partial}_{\alpha}$,
denotes the reduced tridimensional Laplacian (without
$\phi$)
up to a conformal factor: the bidimensional Laplacian is given by
$\Delta ={\partial}^{2}_{\alpha\alpha}$.
Consider now parabolic coordinates, which in terms of Cartesian ones
are given by
\begin{equation}
x=\lambda\mu\cos\phi\;,\;y=\lambda\mu\sin\phi\;,\;
z=\frac{1}{2}({\lambda}^2-{\mu}^2).
\label{C}
\end{equation}
Because we are interested in axisymmetric solutions, we will use polar
cylindrical coordinates $\rho$ and $z$ with
\begin{equation}
\rho=\lambda\mu\;\;\;,\;\;\;z=\frac{1}{2}({\lambda}^2-{\mu}^2).
\label{D}
\end{equation}
The inverse of transformations (\ref{D}) are given by
\begin{equation}
\lambda=\sqrt{z+\sqrt{{\rho}^2+z^2}}\;\;,\;\;
\mu=\sqrt{\sqrt{{\rho}^2+z^2}-z}.\label{E}
\end{equation}
Now, if we write the field equations in parabolic coordinates,
i.e. $x^1=\lambda, x^2=\mu$, then for the operator ${\nabla}^2$ in
these coordinates we have
\begin{equation}
{\nabla}^2={\partial}^{2}_{\mu\mu}+{\partial}^{2}_{\lambda\lambda}+
\frac{1}{\lambda}{\partial}_{\lambda}+\frac{1}{\mu}{\partial}_{\mu}\;.
\label{G}
\end{equation} 
The same operator when expressed in cylindrical coordinates is
\begin{equation}
{\nabla}^2={\partial}^{2}_{\rho\rho}+{\partial}^{2}_{zz}+\frac{1}{\rho}
{\partial}_{\rho}\; . \label{H}
\end{equation}
By comparing the expressions (\ref{G}) and (\ref{H}) it is easy to see
that if we take 
a solution of the equations (\ref{B}) with $f=f(\rho), \Phi=\Phi(\rho)$, 
then also 
$f=f(\lambda), \Phi=\Phi(\lambda)$ or $f=f(\mu), \Phi=\Phi(\mu)$
are solutions. In this way, starting with 
a solution with cylindrical symmetries,
we can obtain a parabolic one that is non
polynomial when expressed in cylindrical coordinates.\\
This paper is devoted to the discussion of the solutions so obtained
together with an investigation of the limits of validity of the method.\\
In section 1 we apply the method using as starting metric the Lewis
\cite{Lew} and the Papapetrou classes \cite{Pap} of solutions 
and discuss possible physical interpretations of these
solutions. In section 2 we show that, starting with a static spatially
homogenous solution with a ${G_3}$ group of motion, a class of stationary
solutions with a ${G_2}$ group of motion can be obtained which contains
as subclass the solution found in subsection 1.1. In section 3 we study
the most general coordinate system permitted by our method. Finally,
section 4 is devoted to a generalization of the method together with a
study of the possible relations with the 
Belinsky-Zakharov (B-Z) solitons-generating solutions.
\section{Application of the method}
\subsection{Generating solutions from Lewis ones}  
For a first application we consider Lewis solutions \cite{Lew}
given by
\begin{eqnarray}
& &f =\frac{1}{(1-B^2)}[P^2{\rho}^{\epsilon}-
B^2 Q^2{\rho}^{2-\epsilon}], \nonumber\\
& &\omega =\frac{B}{PQ}\frac{(Q^2{\rho}^{2-\epsilon}-
P^2{\rho}^{\epsilon})}
{(P^2{\rho}^{\epsilon}-B^2 Q^2{\rho}^{2-\epsilon})},\nonumber\\
& & e^{2\gamma}=\frac{{\rho}^{\frac{{\epsilon}^2-2\epsilon}{2}}}{(1-B^2)}
[P^2{\rho}^{\epsilon}-B^2 Q^2{\rho}^{2-\epsilon}],
\label{N}
\end{eqnarray}
where $B$, $P$, $Q$, and $\epsilon$ are constants.\\
For our purpose, the function $\Phi$  
for the Lewis solutions
must be a function of $\rho$. From equations (\ref{B}) we deduce
\begin{equation}
\Phi=\frac{(\epsilon-1)2BPQz}{{(1-B^2)}}.\label{I}
\end{equation}
Thus, a necessary condition to map cylindrical Lewis solutions into  
parabolic stationary ones is $\epsilon=1$. 
Another possibility is that $B\rightarrow\infty$, but these
solutions belong to the Levi Civita static class 
that will be discussed later.  
Taking $\epsilon=1$ in (\ref{N}) we obtain
$\omega=b$, where $b$ is a real constant.
By integrating the field equations (\ref{B}), we obtain
the solution
\begin{equation}
f=a \lambda\;\;,\;\;\omega=b\;\;,\;\;
e^{2\gamma}=\sqrt{\lambda}{({\lambda}^2+{\mu}^2)}^{\frac{3}{4}},
\label{elbe}
\end{equation}
where $a$ is a real positive constant.
Also the function
\begin{equation}
f=a \mu\label{erere}
\end{equation}
is a solution of the system (\ref{B}) and for $\gamma$ we obtain
the same expression given in (\ref{elbe}) with $\lambda\rightarrow\mu$.
In this way we have used a subclass of Lewis solutions to obtain
two solutions with parabolic-like symmetries.\\
Now we analize the properties of (\ref{elbe}).
First of all, metric (\ref{elbe}) is Petrov type I and has 
an Abelian $G_2$ group of motion with Killing vectors ${\xi}^1={\partial}_t ,
{\xi}^2={\partial}_{\phi}$. 
Besides, it has a coordinate singularity
at $\lambda=0$.
In cylindrical coordinates it takes the form
\begin{eqnarray}
& & ds^2=\frac{a}{2^{\frac{1}{4}}}
\frac{{(z+\sqrt{{\rho}^2+z^2})}^{\frac{-1}{4}}}
{{({\rho}^2+z^2)}^{\frac{1}{8}}}\left[d{\rho}^2+dz^2\right]+
\frac{{\rho}^2}{a}\frac{1}{\sqrt{z+\sqrt{{\rho}^2+z^2}}}
d{\phi}^2-\nonumber\\
& & a\sqrt{z+\sqrt{{\rho}^2+z^2}}
{\left(dt-bd\phi\right)}^2 . \label{c1}
\end{eqnarray}
Since at $\rho=0$ ($z$ axis) 
$e^{2\gamma}=1$ for $z>0$ and $e^{2\gamma}\neq 1$
for $z\leq 0$, we conclude that (\ref{c1}) is regular on the
axis only for $z>0$.\\ 
This fact is confirmed by taking the relativistic 
invariants 
\begin{equation}
R^{abcd}R_{abcd}\;,\;R^{abcd;e}R_{abcd;e}\;,\;
C^{abcd}C_{abcd}\cdots,
\label{nadir}
\end{equation} 
where $R$ denotes the Riemannian tensor, 
$C$ the Weyl one and ``;'' denotes the
covariant derivative. The invariants are coordinate independent
and when expressed in parabolic coordinates they are singular only
for $\lambda=0, {\lambda}^2+{\mu}^2=0$, i.e. on the $z$
axis at $z\leq 0$. For example 
\begin{equation}
R^{abcd}R_{abcd}=\frac{3a^2(4{\lambda}^2+{\mu}^2)}
{4{\lambda}^3{({\lambda}^2+{\mu}^2)}^{\frac{5}{2}}}
\end{equation}
and
\begin{equation}
R^{abcd;e}R_{abcd;e}=\frac{45a^3({\mu}^4+
7{\lambda}^2{\mu}^2+16{\lambda}^4)}{16{({\lambda}^2+{\mu}^2)}^{\frac{17}{4}}
{\lambda}^{\frac{9}{2}}}.
\end{equation} 
To study the behaviour at spatial infinity the spherical coordinates
are most appropriate. When (\ref{elbe}) is expressed
in such coordinates, it is easy to show that the line element is 
not regular
at spatial infinity ($r\rightarrow\infty$) independently of the azimuthal
angle $\theta$, and at $\theta=\pi$ 
($\rho=0, z\leq 0$) independently of $r$.\\
Now we study the physical interpretation of (\ref{elbe}).
As a first step note that it is possible to construct 
(see \cite{Pit} and references therein) a ``local''
formulation of general relativity and thus to define,
in this ``local'' frame,
the analog quantities of the Newtonian theory.
In the stationary case we can define a ``standard'' gravitational
field $G$ in a reference frame $\Gamma$ adapted to the stationary spacetime
(\ref{A})
with a gravitational potential $U$ given by $f=e^{2U}$.\\ 
Besides, in $\Gamma$ we can define a time parameter
$T$ analogue to the time defined in a Galileian
reference frame, except for the fact that $T$ is not defined 
globally but only on the geodesic of the particle.
In terms of the proper time $\tau$, with 
$d\tau=\sqrt{-g_{\alpha\beta}dx^{\alpha}dx^{\beta}}$, we have
\begin{equation}
d\tau = \sqrt{1-{\nu}^2} dT
\label{d1}
\end{equation}
where ${\nu}^{i}=\frac{dx^{i}}{dT}, i=1-3$ is the
3-velocity of a test particle with respect to $T$.\\
Remember that, starting with the line element (\ref{A}), a frame
is adapted to the spacetime (\ref{A}) if it is represented by
coordinates $x^{\prime\; i}=x^{\prime}(x^i),\; 
t^{\prime}=t^{\prime}(x^{\alpha}),\;i=1-3\;\alpha=1-4$.
Further, if $v^{\alpha}$ is the 4-velocity with respect to $\tau$,
then $v^{\alpha}={\nu}^{\alpha}\frac{dT}{d\tau}$.\\ 
In particular,
in this picture we can define the ``relative'' 
energy $H$ of a test
particle as $H=\frac{m_0}{\sqrt{1-{\nu}^2}}e^{U}$
($\frac{dH}{dT}=0$)
, where $m_0$ is
the rest mass of the particle.
For metric (\ref{elbe}), the function $H$ is
\begin{equation}
H=\frac{m_0}{\sqrt{1-{\nu}^2}}\sqrt{f}=
\frac{m_0}{\sqrt{1-{\nu}^2}}\sqrt{a\lambda}.
\label{elika}
\end{equation}
Thus, the surfaces with constant energy are rotational
parabolic. 
Note that if $\lambda\simeq 0$, then $H\simeq 0$. However, it is in principle
possible to have a particle in the orbit with ${\nu}^2=1-a\lambda$
which is therefore ultrarelativistic with $H=m_0$.
Since surfaces with $\lambda=const.$ are equipotential, for
the spacetime (\ref{elbe}) orbits exist with energy
$H=m_0$, i.e. the energy for a rest non-interacting particle.\\
Further, when $\lambda a<1$, the potential
$U=\log\sqrt{f}<0$ and is thus attractive. When
$\lambda=\frac{1}{a}$, $f=1$ and $U=0$: in this case the particle has
``local'' energy $H=\frac{m_0}{\sqrt{1-{\nu}^2}}$ of a free particle
travelling in a Minkowskian spacetime with speed $\nu$.
Finally, for $\lambda>1$ we have $f>1$ and thus the potential $U$
becomes repulsive.\\ 
Naturally, this does not mean that the source matter of
(\ref{elbe}) is a paraboloid, but this
is an indication that parabolic symmetries
have something to do with the solution (\ref{elbe}).
To enforce this reasoning, we consider the coordinate 
transformation \cite{Lew} which changes solution (\ref{elbe})
into the static form
\begin{equation}
ds^2=\frac{e^{2\gamma}}{f}[d{\lambda}^2+d{\mu}^2]+
\frac{{\rho}^2}{f}{d{\phi}^{\prime}}^2-f{dt^{\prime}}^2
\label{c4}
\end{equation} 
with the same functions $\gamma$ and $f$ given in (\ref{elbe}).
This is done by performing the transformation:
\begin{equation}
dt=Adt^{\prime}+Bd{\phi}^{\prime}\;\;,\;\;
d\phi=Cdt^{\prime}+Dd{\phi}^{\prime},
\label{c5}
\end{equation}
where $A,B,C,D$ are functions of the non ignorable coordinates 
$\lambda, \mu$.\\ 
Imposing that (\ref{c4}) be equal to (\ref{elbe}), we 
obtain the set of equations
\begin{eqnarray}
& &\frac{{\rho}^2}{f^2}CD=AB+CD{\omega}^2-AD\omega-BC\omega ,\nonumber\\
& &D^2+\frac{f^2}{{\rho}^2}\left[2BD\omega-{\omega}^2D^2-B^2\right]=1,
\nonumber\\
& &-C^2\frac{{\rho}^2}{f^2}+
A^2+{\omega}^2C^2-2AC\omega=1 .\label{c6}
\end{eqnarray}
System (\ref{c6}) has three independent equations for four variables 
and thus admits  solutions. However, transformations
(\ref{c5}) are purely local, i.e. non integrable. This means
that there exists a rotating reference frame such that the metric
appears to be static, and the coordinates $t^{\prime}, {\phi}^{\prime}$
are admissible only on a geodetic. Further, note that system
(\ref{c6}) depends on $\frac{f^2}{{\rho}^2}$: for the solution
(\ref{elbe}) this quantity is equal to $\frac{a^2}{{\mu}^2}$.
Therefore, the angular velocity of an observer in the frame
$\lambda, \mu, {\phi}^{\prime}, t^{\prime}$ with respect to which
the metric appears to be static, 
depends only on $\mu$ ($\omega$ is constant
in the spacetime (\ref{elbe})).
Thus, the angular velocity of the source,
as ``seen'' from an observer with coordinates 
$\lambda, \mu, \phi, t$,
has a shape
given by a paraboloid of rotation. For example, with the ansatz
$dt=Adt^{\prime}\;,\;d\phi=Cdt^{\prime}+Dd{\phi}^{\prime}$, with
$Q=\frac{d\phi}{dt}$ and $Q^{\prime}=\frac{d{\phi}^{\prime}}{dt^{\prime}}$,
thanks to (\ref{c6}) we obtain
\begin{equation}
Q=\frac{a^2 b}{a^2 b^2-{\mu}^2}+Q^{\prime}\frac{{\mu}^2}{-a^2 b^2+{\mu}^2}.
\label{c7}
\end{equation}
Expression (\ref{c7}) means that the source rotates with an
angular velocity depending on $\mu$. In fact, for 
$Q^{\prime}=0$, it follows that
$Q=\frac{a^2b}{a^2b^2-{\mu}^2}$.
Note that for $\mu\rightarrow\infty$, because in this limit
$dt^{\prime}\rightarrow dt$ and
$d{\phi}^{\prime}\rightarrow d\phi$,
formula (\ref{c7}) reduces to $Q=Q^{\prime}$.
The main difficulty to analize the
nature of the source of (\ref{elbe}) is that this metric does not admit
asymptotical Minkowskian coordinates. Obviously, similar arguments follow
for the solution (\ref{erere}). This concludes our study of the
physical interpretation of (\ref{elbe}).
\subsection{Generating solutions from Papapetrou ones}
\indent Another class of solutions that can be mapped into parabolic
ones is given by Papapetrou \cite{Pap}. 
Starting from the metric (\ref{A}) (Papapetrou gauge), these solutions
are characterized by 
\begin{equation}
f^2+{\Phi}^2=1. \label{P}
\end{equation}
It is easy to see that the most general solution
belonging to the Papapetrou class with
$f=f(\rho), \Phi=\Phi(\rho)$ is 
\begin{equation}
f=\frac{2{\rho}^P}{1+{\rho}^{2P}}\;\;,\;\;
\Phi=\frac{{\rho}^{2P}-1}{{\rho}^{2P}+1},\label{Q}
\end{equation}
where $P$ is a real constant.\\
Once equations (\ref{B}) are solved for the metric functions
$\omega$ and $\gamma$, we obtain 
\begin{equation}
\omega=Pz+\beta\;\;\;\;,\;\;\;e^{2\gamma}={\rho}^{\frac{P^2}{2}}.
\label{Ceka}
\end{equation}
Note that (\ref{Q}) has
cylindrical symmetries only when $\phi=constant$, i.e. for planes
passing through the $z$ axis.
Actually, for our method to be applicable, we only need
solutions with $f=f(\rho), 
\Phi=\Phi(\rho)$: from the field equations (\ref{B}) it is easy to 
see that this implies that 
$\gamma=\gamma(\rho)$ and 
$\omega=\alpha z+\beta$, where $\alpha, \beta$ 
are real constants. Solution (\ref{Q}) is ``mapped'' into 
\begin{equation}
f=\frac{2{\lambda}^P}{1+{\lambda}^{2P}}\;\;,\;\;
\Phi=\frac{{\lambda}^{2P}-1}{{\lambda}^{2P}+1}.
\label{R}
\end{equation}
By integrating the field equations (\ref{B}) we get
\begin{equation}
\omega=\frac{1}{2}P{\mu}^2+\beta\;\;\;,\;\;\;
e^{2\gamma}=\frac{{\lambda}^{\frac{P^2}{2}}}
{{({\lambda}^2+{\mu}^2)}^{\frac{P^2}{4}-1}}.\label{S}
\end{equation}
Also in this case we can obtain another solution by taking
$\lambda\rightarrow\mu$.\\
Generally, (\ref{R})
admits a $G_2$ Abelian group of motion, i.e. 
${\xi}^{1}={\partial}_{\phi}, {\xi}^{2}={\partial}_t$.
Since condition (\ref{P}) is again valid, the solution (\ref{R})
belongs to the Papapetrou class.\\
With arguments similar to the
ones used in subsection 1.1, it can be shown
that the metric (\ref{R}) has a coordinate singularity at $\lambda=0$,  
is regular on the $z$-axis only at
$z>0$ and is singular for $z\leq 0$, i.e. in the limit
$\mu\rightarrow 0$ it follows that
$e^{2\gamma}\rightarrow 1$. In fact, the invariants are singular
for ${\lambda}=0, {\lambda}^2+{\mu}^2=0$. Solution
(\ref{R}) is Petrov type O (flat) for $P=0$,
is Petrov type D for $P=\pm 2$ and otherwise is Petrov type I.
While at spatial infinity ($r\rightarrow\infty$) 
it is not regular. Besides, for
the ``relative'' energy $H$ of a test particle
of inertial mass ${m_0}$ and velocity $\nu$ we get
\begin{equation}
H=\frac{m_0}{\sqrt{1-{\nu}^2}}e^{U}=
\frac{m_0}{\sqrt{1-{\nu}^2}}\frac{\sqrt{2}}
{\sqrt{{\lambda}^P+{\lambda}^{-P}}}
\label{trek}
\end{equation}
and thus, orbits 
with constant energy in a frame $\Gamma$ 
adapted to the stationary metric
(\ref{A}) are again rotational parabolic surfaces.
Then, for $\lambda\simeq 0$, 
$H\simeq\frac{m_0}{\sqrt{1-{\nu}^2}}\sqrt{2}{\lambda}^{\frac{P}{2}}$ 
and for
$\lambda\rightarrow\infty$, $H\simeq\frac{m_0}{\sqrt{1-{\nu}^2}}
\frac{\sqrt{2}}{{\lambda}^{\frac{P}{2}}}$ : in both cases
$H\rightarrow 0$, i.e. $U\rightarrow -\infty$.
Moreover, for spacetime (\ref{R}), $U\leq 0$ with equality  only at
$\lambda =1$: therefore, at $\lambda =1$ 
$H=\frac{m_0}{\sqrt{1-{\nu}^2}}$.\\
Obviously, also in this case, this does not mean that the source matter
of (\ref{R}) is a paraboloid, because the shape of the configuration
depends on the ``match'' between the gravitational and the 
centrifugal force, which is not given a priori. In practice, the fact that
orbits with parabolic symmetries are allowed does not guarantee that
the source is a paraboloid but is a sign in this direction as well as
the fact that the invariants of (\ref{S}) are singular on the axis
at $\lambda=0$. Finally, since 
for (\ref{R}) $\frac{{\rho}^2}{f^2}=F(\lambda,\mu)$
, the arguments that lead to (\ref{c7}) are not valid,
and consequently the
shape of the angular velocity of the source, as ``seen'' from an observer
at rest in a general spacetime point with coordinates
$\lambda, \mu, \phi, t$,
(since the metric is not
asymptotically flat, there does not exist a ``privileged'' Minkowskian
observer at spatial infinity), has not parabolic symmetries. 
\section{Stationary solutions from static ones}
\noindent To apply our
method in the above section we have taken as ``starting'' metric
a subclass of Lewis solutions 
with a $G_3$ group of motion
and a $G_2$ subclass of Papapetrou solutions 
($\alpha\neq 0$). All the generating solutions have 
a $G_2$ Abelian group of motion. In this section
we show that it is possible to obtain a stationary $G_2$ solution
starting with a static $G_3$ solution with Killing
vectors ${\xi}^1={\partial}_t, {\xi}^2={\partial}_{\phi},
{\xi}^3={\partial}_z$. As an example we start with
the most general  
cylindrically symmetrical static solution with
an Abelian $G_3$ group of motion
(found by Levi-Civita \cite{Lev}) that 
can be obtained from Lewis solutions by setting $B=0$ ($\omega=0$) . 
With the same
technique used above we obtain 
\begin{equation}
f=a{\lambda}^{\epsilon}\;\;\;,\;\;\;
e^{2\gamma}=\frac{{\lambda}^{\frac{{\epsilon}^2}{2}}}
{{({\lambda}^2+{\mu}^2)}^{\frac{{\epsilon}^2}{4}-1}},\label{T}
\end{equation}
and the solution with $\lambda\rightarrow\mu$.\\ 
When expressed in terms of $\rho, z$, with the help of (\ref{E}), all
these parabolic solutions have complicated expressions.
Note that, thanks to the analoguey between the Laplacians (\ref{G}) and
(\ref{H}) we can ``map'' all static cylindrical solutions
into parabolic ones.
Solution (\ref{T}) is Petrov type O for $\epsilon =0,2$ and
otherwise is Petrov type I. Furthermore
\begin{equation}
f=a{\lambda}^{\epsilon}\;,\;e^{2\gamma}=
\frac{{\lambda}^{\frac{{\epsilon}^2}{2}}}
{{({\lambda}^2+{\mu}^2)}^{(\frac{{\epsilon}^2}{4}-1)}}\;,\;
\omega=Q=const.
\label{c8}
\end{equation}
is a stationary solution. This solution has similar features
to (\ref{elbe}), i.e. it has a physical coordinates 
independent singularity on the $z$-axis at $z\leq 0$ and is not regular 
at spatial 
infinity. Note that for $\epsilon=1$ the class of solutions 
(\ref{c8}) reduces to (\ref{elbe}).
Further, for the solution (\ref{c8}),
$\frac{{\rho}^2}{f^2}=\frac{{\mu}^2}{a^2}{\lambda}^{(2-2\epsilon)}$ and thus
the angular velocity of the source, in general, has not 
parabolic symmetries.
Now, the ``local'' energy $H$ in the reference
$\Gamma$ is
\begin{equation}
H=\frac{m_0}{\sqrt{1-{\nu}^2}}\sqrt{a}{\lambda}^{\frac{\epsilon}{2}},
\label{mary}
\end{equation} 
and consequently $-\infty<U<+\infty$.
The arguments used for (\ref{elika}) are still valid.\\
We now consider the interesting case of the static
subclass of (\ref{T}) with $\epsilon=2$. This is a flat 
solution except on the $z$ axis at $z\leq 0$ and at spatial infinity
where the metric is not regular. In parabolic coordinates 
we have
\begin{equation}
ds^2=d{\lambda}^2+d{\mu}^2+{\mu}^2 d{\phi}^2-{\lambda}^2 dt^2.
\label{c9}
\end{equation}
Performing the spatial coordinate transformation
\begin{equation}
\tilde{x}=\mu\cos\phi\;,\;\tilde{y}=\mu\sin\phi\;,\;
\tilde{z}=\lambda ,
\label{ewr}
\end{equation}
the metric (\ref{c9}) becomes
\begin{equation}
ds^2=d{\tilde{x}}^2+d{\tilde{y}}^2+d{\tilde{z}}^2-{\tilde{z}}^2dt^2 .
\label{c10}
\end{equation}
The solution admits a $G_4$ group of motion and is isomorphic to the 
the flat static Das \cite{Kra} solution, but with $\tilde{z}>0$, and
therefore covers the Das solution only for
$z>0$. In parabolic coordinates the four Killing vector are
${\xi}^1=(0,\cos\phi,-\frac{\sin\phi}{\mu},0), 
{\xi}^2=(0,\sin\phi,\frac{\cos\phi}{\mu},0),
{\xi}^3=(0,0,1,0), {\xi}^4=(0,0,0,1)$.\\
Also in this case we can generalize solution (\ref{c10}) to a stationary
one obtaining
\begin{equation}
ds^2=d{\lambda}^2+d{\mu}^2+{\mu}^2 d{\phi}^2-
{\lambda}^2{(dt-Qd\phi)}^2
\label{maret}
\end{equation}
where $Q$ is a constant.
\section{Validity of the method}
\noindent In this section we characterize the most general coordinate
transformation for which our method works.
First of all note that equations (\ref{B}) have been   
written starting with the line element
(\ref{A}) (Papapetrou gauge). It is easy to see \cite{Viag}
that the Papapetrou gauge is preserved if and only if
analytical coordinate transformations are considered, i.e.
${x^1}^{\prime}+\imath{x^2}^{\prime}=F(x^1+\imath x^2)$.
Obviously, the imposition of
the Papapetrou gauge does not represent a loss of generality. Therefore
we can restrict our consideration to analytical coordinate transformations
(note that if such coordinate transformations are not used,
equations (\ref{B}) assume a very complicated expression).\\ 
Since the starting point of our method is the analoguey between Laplacians in
cylindrical and parabolic coordinates, it
is natural to ask if there exist some other coordinates $u,v$, 
such that the 
(reduced) Laplacian takes the form
\begin{equation}
{\nabla}^2={\partial}^{2}_{uu}+{\partial}^{2}_{vv}+
\frac{1}{u}{\partial}_u+\frac{1}{v}{\partial}_v\;.
\label{c1e}
\end{equation}
As a first step we consider separable analytical coordinate
transformations:
\begin{equation}
\rho=U(x^1)V(x^2)\;\;\;,\;\;\;z=z(x^1,x^2).
\label{c11}
\end{equation}
Since the analyticity condition ${\rho}_{x^1}=z_{x^2},\;
{\rho}_{x^2}=-z_{x^1}$ must be imposed, we have 
$U_{x^1}V=z_{x^2},\;UV_{x^2}=-z_{x^1}$: these lead to 
\begin{eqnarray}
& &z=-V_{x^2}\int U dx^1 +h(x^2) ,\nonumber\\
& &\frac{U_{x^1 x^1}}{U}=-\frac{V_{x^2 x^2}}{V}.\label{c12}
\end{eqnarray}
System (\ref{c12}) only admits solutions given by
\begin{eqnarray}
& &U=\sinh x^1\;,\;V=\sin x^2\;,\;z=-\cosh x^1\cos x^2 ,\label{c13}\\
& &U=\cosh x^1\;,\;V=\cos x^2\;,\;z=\sinh x^1\sin x^2 ,\label{c14}\\
& &U=e^{x^1}\;,\;V=\sin x^2\;,\;z=-e^{x^1}\cos x^2 ,\label{c15}\\
& &U=x^1\;,\;V=x^2\;,\;z=\frac{1}{2}[{(x^2)}^2-{(x^1)}^2] .\label{c16}
\end{eqnarray}  
Solutions (\ref{c13}) and (\ref{c14}) represent respectively spheroidal 
prolate and oblate coordinates. While (\ref{c15}) represents 
spherical coordinates and (\ref{c16})  
parabolic coordinates. Coordinates (\ref{c13})-(\ref{c16}) are the only 
analytical
coordinates separable with respect to the operator ${\nabla}^2$
(${\nabla}^2={\partial}^{2}_{x^1 x^1}+{\partial}^{2}_{x^2 x^2}+
F(x^1){\partial}_{x^1}+G(x^2){\partial}_{x^2}$). Note that
the parabolic coordinates are the only ones with respect to
which the ``similarity'' reasonings with the cylindrical polar
coordinates are available. However, the
``similarity'' reasoning with respect to the operator
${\nabla}^2$ can be done for another
pair of coordinates.
In fact, if we consider spheroidal prolate 
coordinates $\mu,\theta$ with $\rho=\sinh\mu\sin\theta,
z=\cosh\mu\cos\theta$, the operator ${\nabla}^2$ is
\begin{equation}
{\nabla}^2={\partial}^{2}_{\mu\mu}+{\partial}^{2}_{\theta\theta}+
\frac{\cosh\mu}{\sinh\mu}{\partial}_{\mu}+
\frac{\cos\theta}{\sin\theta}{\partial}_{\theta}.\label{alfa}
\end{equation}
Now, if we take spherical coordinates compatible \cite{Viag} 
with Papapetrou gauge
(\ref{A}), i.e. $\rho=e^{v}\sin\vartheta, z=e^{v}\cos\vartheta$, we have
\begin{equation}
{\nabla}^2={\partial}^{2}_{vv}+{\partial}^{2}_{\vartheta\vartheta}
+{\partial}_v+\frac{\cos\vartheta}{\sin\vartheta}{\partial}_{\vartheta}.
\label{omega}
\end{equation}
Thus, if we have spherical solutions with $f=f(\vartheta), 
\Phi=\Phi(\vartheta)$, we can obtain prolate spheroidal ones with
$\vartheta\rightarrow\theta$. For example, 
we have the stationary solution in spherical coordinates
belonging to the Papapetrou class given by
\begin{equation}
f=\frac{{\sin}^2\vartheta}{1+{\cos}^2\vartheta}\;\;,\;\;
\Phi=\frac{2\cos\vartheta}{1+{\cos}^2\vartheta}\label{rin}.
\end{equation}
From solution (\ref{rin}) we can obtain another one with
$\vartheta\rightarrow\theta$.\\
What happens if we consider coordinate transformations of the form
$\rho = G(x^1,x^2), z=H(x^1,x^2)$ ? It is easy to show that the
analiticity condition $G_{x^1}=H_{x^2}\;,\;G_{x^2}=-H_{x^1}$ leads
inevitably to the condition $G(x^1,x^2)=U(x^1)V(x^2)$, that has been
analysed above.\\ 
As a final consideration note that, since
$f=f(\lambda)$ and $\Phi={\Phi}(\lambda)$, the term 
$\frac{1}{\mu}{\partial}_{\mu}$ in the operator ${\nabla}^2$ 
as expressed by parabolic
coordinates acts trivially on $f(\lambda), {\Phi}(\lambda)$. This
means that if we consider the coordinate transformation given by
$\rho={\lambda}F(\mu)\;,\;z=\frac{a}{2}{\lambda}^2-\frac{F^2}{2a}
\;,\;a=const.$, the method is again applicable, with the inverse given by
\begin{equation}
\lambda=\sqrt{\frac{1}{a}(\sqrt{{\rho}^2+z^2}+z)}\;\;,\;\;
F=\sqrt{a(\sqrt{{\rho}^2+z^2}-z)}.
\label{c17}
\end{equation}
Obviously, the Papapetrou gauge breaks down, but because of the 
independence of $f$ and $\Phi$ on $\mu$ the field equations
for these functions are again of the form given by the first two
equations of (\ref{B}) with 
${\nabla}^2={\partial}^2_{\lambda\lambda}+
\frac{1}{\lambda}{\partial}_{\lambda}$.\\
However, thanks to the second of equations (\ref{c17}), it is easy
to see that the function $F$ represents
the possibility of performing a general coordinate transformation 
for $\mu$ that does not lead to a different solution. No other
possibilities are allowed.
\section{Further improvements and final remarks}
Our starting point has been the line element (\ref{A}) written
in the Papapetrou gauge. However, the most general form 
of the metric for 
a spacetime admitting a two-dimensional Abelian group of isometries
with Killing vectors ${\partial}_t, {\partial}_{\phi}$ is \cite{Kra}
\begin{equation}
ds^2=\frac{1}{f}[e^{2\gamma}({(dx^1)}^2+{(dx^2)}^2)+W^2d{\phi}^2]
-f{(dt-\omega d\phi)}^2
\label{c20}
\end{equation}
where $W=W(x^1, x^2)$ with $W>0$. 
It is easy to see \cite{Lew} that the field equations
imply 
\begin{equation}
\Delta W= W_{x^1 x^1}+W_{x^2 x^2}=0 .
\label{c21}
\end{equation}
Condition (\ref{c21}) means that $W$ can be choosen as a coordinate.
Further, note that the determinant of the 2-metric $g$ spanned by the Killing
vectors ${\partial}_t , {\partial}_{\phi}$ is
\begin{equation}
{(-det||g||)}^{\frac{1}{2}}=W .
\label{c22}
\end{equation}
The function $W$ characterizes a measure of the area of the orbits
\cite{Al} of the isometry group and thus it has a geometrical significance.
The case with $W=x^1=\rho$ has been analysed above. Therefore, we can start
with a given $W=W(\rho , z)=x^1$ satisfying condition (\ref{c21}) 
and with the other coordinate $U$ given by $W_{\rho}=U_z\;,\;
W_z=-U_{\rho}$ and then apply our method. The field equations for this 
new gauge are again given by (\ref{B}) provided that $\rho$ is substituted 
with $x^1$. The similarity
reasoning with the coordinates $W,U$ is possible only with
the coordinates $\alpha, \beta$ given by
\begin{equation}
W=\alpha\beta\;\;\;,\;\;\;U=\frac{1}{2}({\alpha}^2-{\beta}^2).
\label{c24}
\end{equation}
For example, we can take $W=\lambda$ with $\lambda=\alpha\beta\;,\;
\mu=\frac{1}{2}({\alpha}^2-{\beta}^2)$ and use as a starting
metric the solution (\ref{N}) with $\epsilon =1$ and with $\rho$
substituted with $\lambda$. Therefore, by changing $W(\rho, z)$,
we can increase the class of coordinates with respect to which our
method is applicable.\\ 
As a final consideration, we analyze the
possible relations between the method presented in this paper
and the well known \cite{Bel,Zac} Belinsky-Zakharov solitons-generating
technique. This method derives from the application of the
``quantum inverse scattering method'' (QISM) to spacetimes admitting a
two-dimensional Abelian group of isometries acting orthogonally transitively
on  two-dimensional spacelike or timelike orbits.   
For the spacelike case we have two spacelike Killing vectors and
the Ernst equation becomes hyperbolic.
In what follows we restrict our attention to the 
timelike case.\\
The starting point of the B-Z method is the line element
\begin{equation}
ds^2=F(d{\rho}^2+dz^2)+g_{AB}dx^A dx^B\;,\;A,B=3,4.
\label{c30}
\end{equation}
For a given  initial solution $(F_0, g_0)$ of the Einstein equations,
by introducing a 2 x 2 complex matrix $\psi$,
the field equations for $F_0, g_0$ can be reduced to a pair (Lax pair)
of two first order differential equations  
which are the integrability condition for the Ernst equation
with ${(-det||g_0||)}^{\frac{1}{2}}=\alpha\;,\;\Delta\alpha=0$.
By chosing $\alpha$ as one of the coordinates (the determinant of the 
2-metric spanned by Killing vectors), as the other spatial coordinate
$\beta$ we can
take its coniugate: ${\alpha}_{x^1}={\beta}_{x^2}\;,\;
{\alpha}_{x^2}=-{\beta}_{x^2}$. The B-Z method
can generate new solutions $F, g$ by introducing a spectral
parameter $\zeta$ together with the initial condition
$\psi(\zeta=0)=g_0$ (for more details see \cite{Bel,Zac,Al,Jan}).
Further, soliton solutions can be added depending on the
pole singularities of the matrix $\psi(x^1, x^2, \zeta)$
in the complex plane of the spectral parameter $\zeta$.
The poles of the $\psi$ matrix are a certain function ${\mu}_k={\mu}(\rho,z)$
of the coordinates and the integer index $k$ denotes the number of
solitons added to the initial solution. 
In B-Z the pole trajectory, contrary to the original
QISM, are not constant and are given by
\begin{equation}
{\mu}_k={\omega}_k-z\pm{[{({\omega}_k-z)}^2+{\rho}^2]}^{\frac{1}{2}}
\label{c40}
\end{equation}
where the constant ${\omega}_k$
are the origin of the $z$ coordinates.\\
Finally, for the new solution $F,g$ we have again
\begin{equation}
{(-det||g||)}^{\frac{1}{2}}=\alpha.
\label{c41}
\end{equation}
There are formal analogies between the method presented in this paper 
and the B-Z method. First of all, both of them need a starting
metric and the group of isometries must be at least a $G_2$
Abelian group.
Moreover, the area of the orbits
on the isometries group 
(condition (\ref{c41})), is preserved. Our method requires harmonic
coordinates in order to achieve the conservation in form of the
line element and therefore of the field equations. For the B-Z
method the existence of such coordinates $\alpha, \beta$ is a consequence of
the integrability condition of the Ernst equation (see \cite{Al}).
Moreover, in such coordinates the field equation in B-Z involving $\psi$
assume the most simple form. However, these formal analogies are not
sufficient to etablish a direct link between the two methods.
In fact, in spite of the similarity between the pole trajectory formula and
expression (\ref{E}), we have not at our disposal a simple way
to relate a transformed solution in our method with the number
of solitons that can be added in the B-Z method.\\
However, is interesting to note that in both cases, starting
with a spatially homogenous metric with a $G_3$ group of motion,
the resulting solution has a $G_2$ group of motion \cite{Jan}.
We conclude saying that our method is also applicable for spacetimes
admitting two spacelike Killing vectors. In this case the equation
satisfied by $W$ is the wave equation instead of (\ref{c21}).
Furthermore, for the coordinates $W, U$, the
analytical condition is substituted with
\begin{equation}
W_{x^1}=U_{x^2}\;\;,\;\;W_{x^2}=U_{x^1}\;,
\label{der}
\end{equation}implying that 
$W_{x^1 x^1}-W_{x^2 x^2}=0$.

\end{document}